\documentclass[reprint,amsmath,amssymb,aps,superscriptaddress,onecolumn]{revtex4-1}

\usepackage{amssymb}
\usepackage{amsmath}
\usepackage{graphicx,color}
\usepackage{dcolumn}
\usepackage{bm}
\usepackage{epsfig,color}
\usepackage{epstopdf}

\begin{document}

\date{\today}

\title{Pulse solutions of the fractional effective models of the Fermi-Pasta-Ulam lattice with long-range interactions}

\author{Gervais Nazaire Beukam Chendjou}
\affiliation{Pure Physics Laboratory: Group of Nonlinear
Physics and Complex Systems, Department of Physics, Faculty of
Science, University of Douala, Box 24157, Douala, Cameroon}
\affiliation{The Abdus Salam ICTP, Strada Costiera 11, I-34151
Trieste, Italy} \affiliation{SISSA, Via Bonomea 265, I-34136
Trieste, Italy} \affiliation{Istituto dei Sistemi Complessi, Consiglio Nazionale
delle Ricerche, via dei Taurini 19-00185 Roma, Italy}

\author{Jean Pierre Nguenang}
\affiliation{Pure Physics Laboratory: Group of Nonlinear
Physics and Complex Systems, Department of Physics, Faculty of
Science, University of Douala, Box 24157, Douala, Cameroon}
\affiliation{The Abdus Salam ICTP, Strada Costiera 11, I-34151
Trieste, Italy}
\affiliation{SISSA, Via Bonomea 265, I-34136 Trieste, Italy}

\author{Andrea Trombettoni}
\affiliation{SISSA, Via Bonomea 265, I-34136
Trieste, Italy} \affiliation{INFN, Sezione di Trieste, I-34151
Trieste, Italy}\affiliation{CNR-IOM DEMOCRITOS Simulation Center, Via Bonomea 265,
I-34136 Trieste, Italy} 

\author{Thierry Dauxois}
\affiliation{Univ. Lyon, ENS de Lyon, Univ. Claude Bernard, CNRS,
Laboratoire de Physique, F-69342 Lyon, France}

\author{Ramaz Khomeriki}
\affiliation{Department of Physics, Faculty of Exact and Natural
Sciences, Tbilisi State University, 0128 Tbilisi, Georgia}

\author{Stefano Ruffo}
\affiliation{SISSA, Via Bonomea 265, I-34136 Trieste, Italy}
\affiliation{INFN, Sezione di Trieste, I-34151 Trieste, Italy}
\affiliation{Istituto dei Sistemi Complessi, Consiglio Nazionale
delle Ricerche, via Madonna del Piano 10, I-50019 Sesto Fiorentino,
Italy}

\begin{abstract}
  We study analytical solutions of the Fractional Boussinesq Equation (FBE), which is
  an effective model for the Fermi-Pasta-Ulam (FPU)
  one-dimensional lattice with long-range couplings. The couplings decay as a power-law
  with exponent $s$, with $1<s<3$, so that the energy density is finite, but $s$ is small enough to
  observe genuine long-range effects. 
  The analytic solutions are obtained
  by introducing an ansatz for the dependence of the field on space and time. This allows to reduce
  the FBE to an ordinary differential equation, which can be explicitly solved. The solutions are initially localized
  and they delocalize progressively as time evolves. Depending on the value of $s$ the solution is either a pulse (meaning a bump)
  or an anti-pulse (i.e., a hole) on a constant 
  field for $1<s<2$ and $2<s<3$, respectively.  
\end{abstract}

\maketitle

\section{Introduction}\label{sec:I}

The celebrated Fermi-Pasta-Ulam (FPU) model \cite{EFP65,Thierry} 
consists of a one-dimensional lattice of $N$ identical particles interacting through a nearest neighbour nonlinear potential.
The displacement of the particles 
with respect to their equilibrium positions is described by the function of time $u_i(t)$, with $i$ the
label of the lattice site. The problem under investigation was a
central one in the foundations of statistical mechanics: relaxation to microcanonical equilibrium. The lack
of relaxation turned the problem into a ``paradox'', which led to a long series of investigations that were
summarized in 50-th anniversary volumes~\cite{Gallavotti,Chaos}.
The FPU paper also proved to be seminal in nonlinear dynamics, leading to the pioneering
work on {\it solitons} in the Korterveg-de Vries (KdV) equation by Kruskal and Zabusky~\cite{Zabusky1965} 
and to the study of integrability of nonlinear lattices~\cite{Henon,Flaschka}.

Long-range interactions are ubiquitous in nature, the most common examples being gravitational and Coulomb interactions. Recently, long-range interacting systems have been extensively studied for their, often intriguing, statistical
behaviour~\cite{Campa2009,Bouchet2010,Campa2014,Levin2014}. In fact, these systems exhibit inequivalence of ensembles, breakdown of ergodicity, long-lived quasi-stationary states and suppression of chaos, to name just a few anomalous phenomena. Due to the paradigmatic
importance of the FPU model, it is interesting to explore the effect of long-range interactions in this context. Given the presence
in this Special Issue of several papers on non-equilibrium statistical mechanics, we think that the FPU model provides an appropriate arena
to explore the behaviour of systems that persist out-of-equilibrium for long time and depending on the features of the interactions.

In this paper we focus on the specific role played by long-range interactions by analyzing the dynamical properties of the
one-dimensional FPU model appropriately modified to include long-range couplings among particles~\cite{Miloshevich15,Miloshevich17}. 
Particles interact through couplings that decay with a power-law of the distance among them. The exponent of the
power-law is denoted as $s$. This model has been also recently studied in connection with the problem of
anomalous heat conduction in one-dimensional lattices~\cite{Anteneodo,Lepri1,Lepri2}. 

We have shown in a previous paper~\cite{chendjou2018} that, in the continuum 
limit, the model can be {\it effectively} described by a Fractional Boussinesq Equation (FBE) in terms
of the space-time field $u(x,t)$ for small variations of the amplitude. In fact, the presence of long-range couplings
determines the appearance of nonlocal terms in space in the continuum equation, where the  non-locality is
mathematically represented by fractional derivatives. Fractional differential calculus,
i.e. non-integer derivatives and integrals~\cite{Miller1993,Podlubny1999,Kilbas2006}, is playing a very important role in various fields of
science and technology, such as elasticity, water waves, quantum mechanics, signal analysis, control theory, finance, ...

We here deal with the problem of determining the solutions of the FBE introduced in our previous paper~\cite{chendjou2018}.
In order to achieve this target, we introduce an ansatz to express the solution
in terms of a single variable $\eta=|x|^{\rho}-vt$, where $\rho$ is related to the power-law decay exponent $s$.
This generalizes the usual change of frame of the Galilean transformation which gives the ansatz of the usual form $\rm\eta=x-vt$.

We are able to derive an ordinary differential equation in $\eta$ which can be explicitly solved,
providing solutions of the FBE which depend on the parameters of the model, in particular 
on the exponent $s$. Two classes of solutions emerge depending on the value of $s$: {\it i)} {\it pulse} solutions
displaying a ``bump'' over a constant field in the range $1<s<2$; {\it ii)} {\it anti-pulse} solutions
showing a ``hole'' in a constant field in the range $2<s<3$. 

The paper is organized as follows. In Section~\ref{alpha} we introduce the $\alpha$-FPU model 
with long-range interactions and its continuum limit. In Section~\ref{FCTM} we provide details on the
study of the fractional derivatives involved in the FBE. The ansatz used to determine the solutions is presented in 
Section~\ref{ansatz}. We determine the exact solutions of the FBE in Section~\ref{exsol}. Concluding remarks are given in Section~\ref{conclusions}. 


\section{The FPU model with long-range interactions and its continuum limit}
\label{alpha}

The FPU model we consider is a one-dimensional lattice of $N$ anharmonic oscillators,
each with mass $m$, displacement $u_i$ and momentum $p_i$ ($i$ denoting the lattice site) and coupled by long-range interactions.
The Hamiltonian for the long-range FPU model reads
\begin{equation}
  {\cal H}\left(u_i,\,p_i \right) = \sum_{i}
  { \frac{p_i^2}{2m}  }   +
  \frac{\chi}{2}\sum_{j<i}
       {\frac{{\left( {u_i  - u_j } \right)^2 }}{{\left| {a\left( {i - j} \right)} \right|^{s} }}}
       + \frac{\gamma}{3}\sum_{j<i}
       {\frac{{\left( {u_i  - u_j } \right)^{3}}}{{\left| {a\left( {i - j} \right)} \right|^{s} }}}.
       \label{alpha2}
\end{equation}
We can observe that the two-body coupling is assumed to be translationally invariant since it acts between pairs of sites at
distance $a|i-j|$. More importantly for our purposes, the coupling decays for large distances as a power-law with an exponent $s$. 
In order for the energy to be extensive, we consider $s$ larger than the dimension $d$ where the system is embedded, in this case $d=1$.
The lattice sites $i$ and $j$ take all possible negative and positive integer values: $i,j=0,\pm 1, \pm 2, \cdots$. We consider
$1<s<3$ since at $s=3$ the system is supposed to have a typical short-range long-time behaviour~\cite{chendjou2018}. The fact
that it exists a value of $s$, denoted by $s^*$, such that
for $s>s^*$ one retrieves the critical properties of the short-range case ($s \to \infty$) is well studied
in Ising and $O(n)$ models~\cite{Defenu2015,Defenu2016}. Notice that the range $d<s<s^*$ is often referred to as the weak
long-range region \cite{Defenu19}. In Eq.~(\ref{alpha2}) we assume also that both the harmonic and anharmonic couplings
decay with the same exponent. One could study as well the case in which the two power-law decay exponents are different~\cite{chendjou2018}. 
Finally, we observe that, since the anharmonic term is characterized by the power $3$, the resulting FPU model is called $\alpha$-FPU~\cite{Livi85}, so that Eq.~(\ref{alpha2}) corresponds to the Hamiltonian of the long-range $\alpha$-FPU model.

The interaction potential energy has two parts: the second term of the right in Eq.~(\ref{alpha2}) is the long-range harmonic interaction,
whereas, the third term corresponds to the long-range anharmonic interaction.
The parameters $\chi$ and
$\gamma$ measure the strength of the linearity and nonlinearity, respectively,
and they can be viewed as the linear and nonlinear spring coefficients, respectively. In the following we consider the case of positive parameters
$\chi,\gamma>0$.
The strength of the long-range interaction decays as a power-law $\left|a( i-j )\right|^{-s}$
where $a$ is the lattice spacing.
The parameter $s$ sets then the range of the interaction. For $s \to \infty$, the second and the third terms reduce
to nearest neighbour, short-range, harmonic and anharmonic interactions, respectively, and we return to standard short-range
FPU model.

In order to study the dynamical properties we move to the effective model describing the continuum limit of the FPU
Hamiltonian~(\ref{alpha2}), where $u_i(t) \to u(x,t)$. This effective model was derived in \cite{chendjou2018}
and the corresponding equation for the field $u(x,t)$ is a fractional differential equation, the FBE, which generalizes
to the weak long-range region the Bousinessq equation found for the short-range FPU model \cite{Davydov1980}. We refer
to \cite{chendjou2018} for a discussion of the details needed to perform the continuum limit and its range of validity, which essentially
requires to consider a slowly varying in space field $u$. We also point out that the long-range FPU model has the merit to
allow for a microscopic derivation of the fractional differential equation, which is not -- as done in other cases -- assumed on
the basis of physical considerations.

The FBE for the long-range $\alpha$-FPU model reads \cite{chendjou2018}
\begin{equation}
\frac{{\partial ^2 u\left( {x,t} \right)}} {{\partial t^2 }} \,\,-\,\,
g_{s - 1} \frac{{\partial ^{s - 1} u\left( {x,t} \right)}}
{{\partial \left| x \right|^{s - 1} }} \,\,-\,\, h_{s - 1} u\left(
{x,t} \right)\left[ {D_{x^ -  }^{s - 1}  - D_{x^ +  }^{s - 1}
} \right]u\left( {x,t} \right) = 0
 \label{alpha3},
\end{equation}
where the constants $g_{s-1} $ and $h_{s-1}$ are given by
\begin{equation}
g_{{s} - 1}  = \frac{{\chi \pi }}{{\Gamma \left({s}\right)\sin
\left( {\frac{{{s} - 1}}{2}\pi} \right)}}, \,\,\,\,\,\, h_{{s} - 1}  = -\frac{{\gamma \pi }}{{\Gamma \left({s}\right)\sin
\left( s\pi \right)}}
\label{alpha4}
\end{equation}
($\Gamma$ stands for the Euler-gamma function, $m$ and $a$ are constants of order unity). In the following we are going to take $s \neq 2$.

Eq.~(\ref{alpha3}) is the equation for which analytical solutions are needed to be found,
which is the task we deal with in the next Sections. The analytical solution then provide a guide to the numerical solution
of the FPU lattice model, studied in the regime in which the continuum limit is a reasonable approximation. Of course, such a comparison
is a benchmark for the assessment of the validity of the effective model. Moreover, as we point out in the following, the solutions
of the effective model may reveal new class of solutions,
possibly not previously known, and motivate further numerical studies
of the lattice equations.

\section{The fractional chain rule}
\label{FCTM}

In order to construct exact solutions, we need to give some details on the action of the fractional derivatives appearing
in Eq.~(\ref{alpha3}).

We first observe that the chain rule can be written as 
\begin{widetext}
\begin{equation}
\frac{{\partial ^\alpha \, w}}
{{\partial x^\alpha  }} = \sigma_\alpha\, \frac{{\partial ^\alpha  \zeta }}
{{\partial x^\alpha  }}\,\frac{{\partial \,w}}
{{\partial \zeta }}
\label{alpha02},
\end{equation}
\end{widetext}	
in general valid for analytic functions of the form $w\left(x\right)=\sum_{m=0}^{\infty}{a_m \,x^m}$,
where $a_m$ are constants. $\sigma_\alpha$ is sometimes called fractal index \cite{Tarasov2016} and it has to to be further determined.
It is usually determined in terms of  Euler-gamma functions. The reason for studying the relation~(\ref{alpha02}) is that
it can be used to transform the FBE into an ordinary differential equation.

To show the validity of~(\ref{alpha02}), we may use the function $w(\zeta)=\zeta^\beta$ with $\zeta(x)=x^\gamma$ where $\beta, \,\, \gamma >0$,
and the definition of the fractional derivative of non-integer order which reads\begin{equation}
  \frac{\partial^\alpha}{\partial x_{+}^\alpha}x^\beta  \equiv \frac{\partial^\alpha}{\partial x^\alpha}x^\beta = \frac{\Gamma \left(1+ \beta \right)}{\Gamma\left(1+ \beta - \alpha \right)}x^{\beta-\alpha }\,\,\,\,\,\,\,\,\,\,\,\,\, x>0,  \,\,\, and \,\,\,\,\,\,\, \alpha>0
 \label{alpha03}.
\end{equation}

One gets
\begin{equation}
D_{x}^\alpha\, w(\zeta(x))= D_{x}^\alpha x^{\beta\gamma}= \frac{\Gamma \left(1+ \beta\gamma \right)}{\Gamma\left(1+\beta \gamma - \alpha \right)}\,\,x^{\beta\gamma-\alpha }
 \label{FCT1},
\end{equation}
\begin{equation}
D_{x}^\alpha {\zeta(x)} \equiv \frac{{\partial ^\alpha \zeta  }}
{{\partial x^\alpha  }}= \frac{{\partial ^\alpha x^{\gamma}  }}
{{\partial x^\alpha  }} = \frac{\Gamma \left(1+ \gamma \right)}{\Gamma\left(1+\gamma - \alpha \right)}\,\,x^{\gamma-\alpha }
 \label{FCT2},
\end{equation}
and 
\begin{equation}
\frac{{\partial \,w}}{{\partial \zeta }} =\left (\beta\, \zeta^{\beta-1}\right){|}_{\zeta=x^\gamma}\,\,=\beta\, x^{{(\beta-1)}\gamma}
 \label{FCT3}.
\end{equation}

Substitution of  Eqs.~(\ref{FCT1}), (\ref{FCT2}) and (\ref{FCT3}) into  Eq.~(\ref{alpha02}) gives
\begin{equation}
\left(  \frac{\Gamma \left(1+ \beta\gamma \right)}{\Gamma\left(1+\beta \gamma - \alpha \right)} -    \sigma_\alpha\, \frac{\beta\,\,\Gamma \left(1+ \gamma \right)}{\Gamma\left(1+\gamma - \alpha \right)}  \right)x^{\beta\gamma-\alpha }=0, \,\,\,\,\,\,\,\,\, x>0
 \label{FCT4}.
\end{equation}

As a result, we have the condition
\begin{equation}
\frac{\Gamma \left(1+ \beta\gamma \right)}{\Gamma\left(1+\beta \gamma - \alpha \right)} -    \sigma_\alpha\, \frac{\beta\,\,\Gamma \left(1+ \gamma \right)}{\Gamma\left(1+\gamma - \alpha \right)} =0, \,\,\,\,\,\,\, \alpha, \beta, \gamma>0
 \label{FCT5}.
\end{equation}
Assuming 
\begin{equation}
 \sigma_{\alpha}= \frac{{\Gamma\left(1+\beta \gamma \right)}{\Gamma\left(1+ \gamma - \alpha \right)}}{\beta\,{\Gamma\left(1+\beta \gamma - \alpha \right)}{\Gamma\left(1+ \gamma  \right)}}
 \label{FCT6},
\end{equation}
one immediately sees that Eq.~(\ref{alpha02}) holds.
We then conclude that the modified chain rule given by
Eq.~(\ref{alpha02}) holds for the correct choice of the fractal
index $\sigma_\alpha$.
We emphasize that the fractal index has to be determined separately
in each problem.  

Since
\begin{equation}
\frac{\partial^\alpha}{\partial  \left| x \right|^\alpha} \equiv -\frac{1}{2\cos\left(\frac{\alpha \pi}{2} \right)}\left[ D_{x^-}^\alpha  \,\,+ \,\,D_{x^+}^\alpha \right], \,\,\,\,\,\,\,\,\,\,\,\,\,\,\,\,\,  \,\,\,\,\,\,\,\,\,\,\,\,\,\,\,\,\, i.e., \,\,\,\,\,\,\,\,\,\,\,\,\,\,\,\,\,  \,\,\,\,\,\,\,\,\,\,\,\,\,\,\,\,\,  D_{x^-}^\alpha  \equiv -\,2\,\cos\left(\frac{\alpha \pi}{2} \right)\frac{\partial^\alpha}{\partial  \left| x \right|^\alpha}\,\, -\,\, D_{x^+}^\alpha
\label{alpha5},
\end{equation}
Eq.~(\ref{alpha3}) can be rewritten as follows:
\begin{equation}
\frac{{\partial ^2 u\left( {x,t} \right)}} {{\partial t^2 }} \,\,\,-\,\,\,
g_{s - 1} \frac{{\partial ^{s - 1} u\left( {x,t} \right)}}
{{\partial \left| x \right|^{s - 1} }} \,\,\,-\,\,\, q_{s - 1} u\left(
{x,t} \right)\frac{{\partial ^{s - 1} u\left( {x,t} \right)}}
{{\partial \left| x \right|^{s - 1} }}\,\,\,+\,\,\, h_{s - 1} u\left(
{x,t} \right)\frac{{\partial ^{s - 1} u\left( {x,t} \right)}}
{{\partial x ^{s - 1} }}\,\, = \,\,0
 \label{alpha6},
\end{equation}
where 
\begin{equation}
q_{{s} - 1}  = \frac{{\gamma\pi }}{{\Gamma \left({s}\right)\cos
\left( {\frac{{{s} }}{2}\pi} \right)}}
\label{alpha06}.
\end{equation}

\section{Ansatz for the dynamics}
\label{ansatz}

A possible way to construct solutions of Eq.~(\ref{alpha6}), the one we explore in this paper, is to assume that the field
$u(x,t)$ depends not on $x$ and $t$ separately, but via a combination of them defining a new variable $\eta$. That is
typical when looking for solutions having solitonic form, and in that case one puts $\rm{\eta = x -vt}$, as routinely done in studying
soliton solutions of nonlinear differential equations \cite{Pitaevskii}. However, despite one can certainly investigate
such dependence, putting $\rm{\eta=x-vt}$ does not reduce the FBE to an ordinary differential equation. By inspection one sees that this a direct
consequence of the fractional derivative, and that in turn it is caused by the presence of the long-range terms. Moreover,
one has a derivative with respect to the modulus of the variable $x$, also in the linear case ($\gamma=0$). Therefore,
one is lead to consider the following ansatz:
\begin{equation}
G\left(\eta\right) \equiv u\left(x,t \right), \,\,\,\,\,\,\,\,\,\,\,\,\,\,\,\,\,\,\,\,\,\,\,\,\,\,\,\,\,\,{\rm{where}}\,\,\,\,\,\,\,\,\,\,\,\,\,\,\,\,\,\,\,\,\,\,\,\,\,\,\,\,\,\,
\eta  =\left| x \right|^{\rho}- v\,t
 \label{alpha77},
\end{equation}
with $v$ a parameter having the dimension $[v]\,=\,L^{{}^{\Huge{\rho}}}\, T^{-1}$ and $\rho$ to be fixed. The ansatz~(\ref{alpha77}) is often referred to as the fractional complex transform in the context of
fractional differential equations \cite{He2010,Tarasov2013}.
Notice that with the ansatz~(\ref{alpha77})
the cone-like structure in the $(x-t)$ space defined by $\eta=0$ is no longer a straight line, and this corresponds
to accelerating or decelerating fields $u$ according the value of $\rho$. One readily observes that to reduce the
FBE to an ordinary differential equation one needs to put
\begin{equation}
\rho \equiv s-1
 \label{alpha_bis}.
\end{equation}

We pause here to comment that the ansatz~(\ref{alpha77})
is less innocent that it may at first sight appears. First, it
breaks the Galilean
invariance, which is broken evidently by the initial condition. Second, it involves the space variable $x$ in the form
$|x|$, so that this implies that care has to be put on choosing the matching conditions in $x=0$. Moreover, to compare
with numerical simulations on the lattice one has to choose even initial conditions for the $u$. Third, and more important, 
by considering the ansatz~(\ref{alpha77}) one is also assuming
that if the $u(x,t=0)$ at the initial time is not vanishing, also its velocity $\dot{u}(x,t=0)$ is not vanishing. In other words,
the ansatz~(\ref{alpha77}) corresponds to a class of solutions with certain conditions to be satisfied at the initial time $t=0$. 

To further proceed, we observe that it exists a duality between left and right fractional derivatives \cite{Caputo2014}:
\begin{center}
$D_{x^{+}}^\alpha f\left( {x} \right)=D_{x^{-}}^\alpha f^{*}\left( {-x} \right ) \,\,\,\,with\,\,\,\,\,f^{*}\left( {x} \right )=f\left( {-x} \right )$.
\end{center}
Then it follows
\begin{equation}
\frac{\partial^\alpha}{\partial x_{-}^\alpha}(-x)^\beta \equiv \frac{\partial^\alpha}{\partial (-x)^\alpha}(-x)^\beta = \frac{\Gamma \left(1+ \beta \right)}{\Gamma\left(1+ \beta - \alpha \right)}(-x)^{\beta-\alpha }\,\,\,\,\,\,\,\,\,\,\,\,\,\,\,\,\,\,\,\,\,\,\,\,\,\,\,\,\,\,\,\,\,\,\,\,\,\,\,\,\,\,\, \,\,\,\,\,\,\,\,\,\,\,\,\,\,\,\,\,\,\,\,\,\,\,\,\,\,\,\,\,\,\,\,\,\,\,\,\,\,x<0
 \label{alpha08},
\end{equation}
and 
\begin{equation}
\frac{{\partial ^{\alpha} }}{{\partial \left| x \right|^{\alpha} }}\left| x \right|^{\beta} = -\, \frac{\Gamma\left( \beta +1 \right)}{\Gamma\left( \beta +1 - \alpha \right) \cos{(\frac{\alpha}{2}\pi)}}\,\left| x \right|^{\beta-\alpha}, \,\,\,\,\,\,\,\,\,\,\,\,\,\,\,\,\,\,\,\,\,\,\,\,\,\,\,\,\,\,\,\,\,\,\,\,\,\,\,\,\,\,\,\,\,\,\,\,\,\,\,\,\,\,\,\,\,\,\,\, \alpha \ne 2\,k\,+\,1, \,\,\,\,\,  k \in \mathbb{R}
 \label{alpha008}.
\end{equation}

In the case $\alpha \to 2$, corresponding to $s \to 3$,
we retrieve the well known integer classical derivative.
Using Eqs.~(\ref{alpha02}), (\ref{alpha03}), (\ref{FCT6}), (\ref{alpha77})
and (\ref{alpha008}) into  Eq.~(\ref{alpha6}) we obtain the following ODE:
\begin{equation}
 v^2\,{\frac{{d^2 G(\eta)}}{{d\eta ^2 }}\,\,\,\, +\,\, \,\,\tilde g_{s }\,\frac{{d G(\eta)}} {{d\eta }}\,\, \,\,+\,\,\,\, \tilde h_{s }\,G(\eta)\,\frac{{d G(\eta)}}{{d\eta }} = 0} ,
 \label{alpha9}
\end{equation}	
where $$\tilde g_{s } \equiv -\frac{2\,\chi\,\pi}{\Gamma^2\left( s \right)\,\sin\left( s \pi \right)}$$ and
$$\tilde h_{s } \equiv \frac{\gamma\,\pi}{\Gamma^2\left( s \right)\,\sin\left( s \pi \right)}.$$

Integrating Eq.~(\ref{alpha9}) with respect to $\eta$ yields
\begin{equation}
v^2\,\frac{{d G(\eta)}}{{d\eta }}\,\,   +\,\, { \tilde g_{s}\, G(\eta)\,\, +\,\, \frac{{ \tilde h_{s}
}}{2}\, {G^2(\eta)} }  \,\,+ \,\,\delta_0\, =\,0
\label{alpha10},
\end{equation}
where $\delta_0$ is a constant to be later determined using appropriate boundary conditions. From Eq.~(\ref{alpha10}) we can write
\begin{equation}
 \int\limits {\frac{{dG}}{{{ \delta_0  \,\,+\,\,\tilde g_{s}\,G \,\,+ \,\,\frac{{ \tilde h_{s }
}}{2}\,G^2 } }}} = -\frac{1}{v^2}\,\int\limits { d\,\eta  } 
\label{alpha12}.
\end{equation}
It is 
\begin{equation}
\begin{array}{*{20}c}
 \int\limits {\frac{{dX}}{{{ a \,\,+ b\,X }  + c \,X^2  }}}= \left\{ {\begin{array}{*{20}c}
-\frac{2}{\sqrt{-\Delta}}arc\tanh{\left(\frac{b\,+2\,c \,X}{\sqrt{-\Delta}}\right)} \,\,\,\,\,\,\,\,\,\,\, \,\,\,\,\,\, \rm{for} \,\,\,\,\,\,\,\,\,\,\,\,\,\,\,\,\,\, \Delta<0, \\
  \\
  -\frac{2}{b\,+\,c\,X}\,\,\,\, \,\,\,\,\,\, \,\,\,\,\,\,\,\,\,\,\, \,\,\,\,\,\,  \,\,\,\,\,\,\,\,\,\,\, \,\,\,\,\,\, \,\,\,\,\,\,\,\,\,\,\, \,\,\,\,\,\,\,\, \rm{for} \,\,\,\,\,\,\,\,\,\,\,\,\,\,\,\,\,\, \Delta=0, \\
  \\
  \frac{2}{\sqrt{\Delta}}arc\tan{\left(\frac{b\,+2\,c \,X}{\sqrt{\Delta}}\right)} \,\,\,\, \,\,\,\,\,\,\,\,\,\,\,\,\,\,\,\,\, \,\,\,\,\,\,
  \rm{for} \,\,\,\,\,\,\,\,\,\,\,\,\,\,\,\,\,\, \Delta>0, \\
 \end{array} } \right.
  \\
 \end{array}
 \label{alpha13}
\end{equation}
with $\Delta \equiv 4\,a\,c\,-\,b^2$ \cite{Gradshteyn2007}.

\section{Solutions of the Fractional Bousinessq Equation}
\label{exsol}

Using the results of the previous Section, via the identification
$a=\delta_0$, $b=\tilde g_s$ and $c=\tilde h_s/2$, one finds that 
the considered solution of Eq.~(\ref{alpha9}) reads
\begin{equation}
\begin{array}{*{20}c}
G\left( \eta \right)=  \left\{ {\begin{array}{*{20}c}
2\,\frac{\chi}{\gamma} \,\,+\,\,\frac{\sqrt{-\Delta}}{\tilde h_s}\,\,\tanh{\left[ \frac{\sqrt{-\Delta}}{2\,v^2}\,\left(\eta \,-\, \eta_0 \right)  \right]} \,\,\,\,\,\, \,\,\,\, \,\,\,\,\,\,\,\,\,\,\,\,\,\,\,\,\, \,\, \Delta <0,\\
  \\
 -\, \frac{4\,\chi}{\gamma} \,\,+ \,\,\frac{4\,v^2}{\tilde h_s \, \left(  \eta \,-\, \eta_0 \right)}\,\,\,\,\,\,\,\,\,\,\,  \,\,\,\, \,\,\,\,\,\,\,\,\,\,\,\,\,\,\,\,\, \,\,\,\,\,\,\,\,\,\,\, \,\,\,\, \,\,\,\,\,\,\,\,\,\,\,\,\,\,\,\,\, \,\,\,\,\,\,\,\,\,\,\,  \Delta =0,\\
  \\
2\,\frac{\,\chi}{\gamma} \,\,-\,\,\frac{\sqrt{\Delta}}{\tilde h_s}\,\,\tan{\left[ \frac{\sqrt{\Delta}}{2\,v^2}\,\left(\eta \,-\, \eta_0 \right)  \right]} \,\,\,\,\,\,\,\,\,\,\,\,\,\,\, \,\,\,\, \,\,\,\,\,\,\,\,\,\,\,\,\,\,\,\,\, \,\, \Delta >0,\\
 \end{array} } \right.
  \\
 \end{array}
 \label{alpha14}
\end{equation}
where
\begin{equation}
\Delta=2\delta_0 \tilde h_s - \tilde g_s^2=\frac{2\,\pi\,\gamma\,}{ \Gamma^2\left( s \right)\sin\left( s \pi \right)}\left[ \delta_0  - \frac{2\,(\chi^2/\gamma)\pi}{\Gamma^2\left( s \right)\sin\left( s \pi \right)} \right]
\label{alpha15}.
\end{equation}
Without any loss of generality we can make the choice $\eta_0=0$.

We have now to fix the boundary conditions. We look for the solutions
such that the following boundary conditions are satisfied: 
 \begin{equation}
\left\{ {\begin{array}{*{20}c}
G\left( \eta = 0 \right)=A,\\
  \\
G \left( \eta \to +\infty \right)=C,\\
\end{array}} \right.
 \label{alpha09}
\end{equation}	
where $A$ and $C$ are two constants to be later specified. Combining Eqs.~(\ref{alpha14}) and (\ref{alpha09}), it turns out that the solution reads
\begin{equation}
G\left( \eta \right)= 2\,\frac{\chi}{\gamma}\,\, +  \,\,\frac{\sqrt{-\Delta}}{\tilde h_s}\tanh{\left( \frac{\sqrt{-\Delta}}{2\,v^2}\,\eta \right)}
\label{alpha16}.
\end{equation}
Eq.(\ref{alpha16}) can be written also as
\begin{equation}
u\left( x,t \right)= 2\,\frac{\chi}{\gamma}\,\, +  \,\,\frac{\sqrt{-\Delta}}{\tilde h_s}\tanh{\left[ \frac{\sqrt{-\Delta}}{2\,v^2} \left( \left|x\right|^{s-1} -v\,t  \right) \right]}
\label{alpha17},
\end{equation}
where we have to specify the parameter $\delta_0$ in terms of $A$ and $C$.
One sees that
$$
A \equiv 2\,\frac{\chi}{\gamma},
$$
with the constant $A$ to be taken positive, in agreement with our choice
of considering $\chi$ and $\gamma$
positive parameters. Moreover it has to be 
$$
C \ne A.
$$
The point essential for the subsequent discussion is that
$$C=A+\frac{\sqrt{-\Delta}}{\tilde h_s}.$$ Since for
$1<s<2$ one has $\tilde h_s<0$ and for $2<s<3$ simply implies that 
it is $\tilde h_s>0$, it follows that $C<A$ for $1<s<2$
and  $C>A$ for $2<s<3$

Given $C$, $\Delta$ is given by
$$
\Delta=-\tilde h_s^2 \left( C-A \right)^2<0,
$$
so that the first choice in the top of Eqs.~(\ref{alpha14}) is justified.
From this expression for $\Delta$ one gets the following value for $\delta_0$:
\begin{equation}
\delta_0 = - \frac{\tilde h_s^2 \left( C-A \right)^2 - \tilde g_s^2}{2 \tilde h_s}.
 \label{alpha18}
\end{equation}

\begin{figure}[h!]
\centering
\includegraphics[width=4.0in,angle=0]{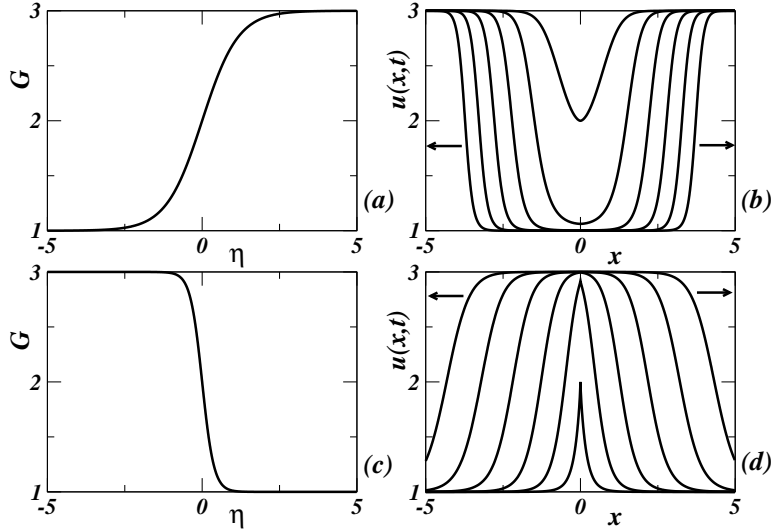}
\caption{Panels (a)-(b): $s=2.75$, with $G(\eta)$ vs $\eta$ in (a)
  and $u(x,t)$ for different times $t=0,2,4,6,8,10$. The values
  of $A$ and $|C-A|$ are $2$ and $1$, respectively.
  Panels (c)-(d): $s=1.75$, with $G(\eta)$ vs $\eta$ in (c)
  and $u(x,t)$ for different times $t=0,0.6,1.2,1.8,2.4,3$ (again with
  $A=2$ and $|C-A|=1$). $v$, $\chi$ and $\gamma$ are constants of order unity.
  }
\label{solutions 1}
\end{figure}

The behaviour of the function $G$ and therefore of $u(x,t)$ is summarized in Fig.~\ref{solutions 1}. 
Panels Fig.~\ref{solutions 1}(a)-(b) refers to a value of $s$ between $2$ and $3$. The $G(\eta)$ is 
increasing for $\eta$ passing from negative values to positive ones. In this case $A$ and $C$ has to be 
both positive, with $C>A$. The initial condition $u(x,t=0)$ has a ``hole'', to which we may refer as
an anti-pulse (notice that $G$ has a typical kink-like form). As soon as that $t$ increases the 
field $u(x,t)$ tends to reach the value $G(\eta \to -\infty)$, i.e., $A-|C-A|$, showing the delocalization 
of the initial condition. It is interesting to observe that the values of $x$ and $t$, denoted by
$\bar{x}$ and $\bar{t}$ such that $u(\bar{x},\bar{t})=A$ are defined by the relation $\bar{\eta}=0$, i.e.
$|\bar{x}|^{s-1}=v\bar{t}$. Since $d\bar{x}/d\bar{t}=[v^{1/(s-1)}/(s-1)]\bar{t}^{\left(1/(s-1)-1\right)}$,
for $2<s<3$ one sees that the velocity decreases and go to zero for large time, so that the propagation
of the hole {\em decelerates}. At variance panels Fig.~\ref{solutions 1}(c)-(d) refers to a value of $s$ 
between $1$ and $2$. The $G(\eta)$ is now decreasing  for $\eta$ passing from negative values to positive
ones. In this case $A>0$ and $C<A$. The initial condition $u(x,t=0)$ has a ``bump'', to which we may refer
as a pulse (or an anti-kink for the $G$). When $t$ increases the field $u(x,t)$ tends to reach the value 
$G(\eta \to +\infty)$, i.e., $A+|C-A|$, showing also now the delocalization of the initial condition.
Since $d\bar{x}/d\bar{t} \propto \bar{t}^{\left(1/(s-1)-1\right)}$, for $1<s<2$ one sees that the velocity increases so that 
the propagation of the bump {\em accelerates} [reaching infinite velocity for $s=1$, which represents the mean-field limit].
\vspace{0.35cm}

To conclude, we have to show that the results above does not qualitatively
depend on the choice of the boundary condition~(\ref{alpha09}). Suppose indeed
that one fixes the function $G$ in two points $\eta_1, \eta_2$, with the
boundary conditions~(\ref{alpha09}) corresponding to $\eta_2=0$ and
$\eta_1 \to \infty$. Using~(\ref{alpha16}) one has
\begin{equation}
G\left( \eta_1 \right)\,-\,G\left( \eta_2 \right)\,= \,\frac{\sqrt{-\Delta}}{\tilde h_s}\left[\tanh{\left( \frac{\sqrt{-\Delta}}{2\,v^2}\,\eta_1 \right)}\,-\,\tanh{\left( \frac{\sqrt{-\Delta}}{2\,v^2}\,\eta_2 \right)}\right]
\label{app_2}.
\end{equation}
Since $\tanh{\left( a-b \right)}=\frac{\tanh{\left( a \right)}\,-\,\tanh{\left( b \right)}}{1\,-\,\tanh{\left( a\right)}\tanh{\left( b \right)}}$,
one can rewrite Eq.~(\ref{app_2}) as follows:
\begin{equation}
G\left( \eta_1 \right)\,-\,G\left( \eta_2 \right)\,= \,\frac{\sqrt{-\Delta}}{\tilde h_s}\tanh{\left( \frac{\sqrt{-\Delta}}{2\,v^2}\,\left(\eta_1\,-\,\eta_2\right) \right)}\left[1\,-\,\tanh{\left( \frac{\sqrt{-\Delta}}{2\,v^2}\,\eta_1 \right)}\tanh{\left( \frac{\sqrt{-\Delta}}{2\,v^2}\,\eta_2 \right)}\right]
\label{App_3}.
\end{equation}
Assuming $\eta_1>\eta_2>0$ one has
$0<\tanh{\left( \frac{\sqrt{-\Delta}}{2\,v^2}\,\eta_1 \right)}\tanh{\left( \frac{\sqrt{-\Delta}}{2\,v^2}\,\eta_2 \right)}\le 1$, i.e.,
\begin{equation}
1\,-\,\tanh{\left( \frac{\sqrt{-\Delta}}{2\,v^2}\,\eta_1 \right)}\tanh{\left( \frac{\sqrt{-\Delta}}{2\,v^2}\,\eta_2 \right)}\ge 0
\label{app_5}.
\end{equation}
Eq.~(\ref{app_5}) simply implies that the sign of
$G\left( \eta_1 \right)\,-\,G\left( \eta_2 \right)$ depends of the sign of
$\frac{\sqrt{-\Delta}}{\tilde h_s}\tanh{\left( \frac{\sqrt{-\Delta}}{2\,v^2}\,\left(\eta_1\,-\,\eta_2\right) \right)}$. With $\eta_1 > \eta_2 >0$ as we
are assuming, it is 
\begin{equation}
\tanh{\left( \frac{\sqrt{-\Delta}}{2\,v^2}\,\left(\eta_1\,-\,\eta_2\right) \right)} >0
\label{app_6}.
\end{equation}
One then sees that the sign of $G\left( \eta_1 \right)\,-\,
G\left( \eta_2 \right)$ depends only on the sign of $\tilde h_s$:
since it is $\tilde{h}_s<0$ for $1<s<2$ and $\tilde{h}_s>0$  for $2<s<3$ we conclude that indeed
\begin{equation}
\left\{ {\begin{array}{*{20}c}
    \,\,\,\,\,\,\,\,\,\,G\left( \eta_1 \right)\,-\,G\left( \eta_2 \right)<0,\,\,\,\,\,\,\,\,\, \text{for} \,\,\,\,\,1<s<2,\,\,\,\,\,\,\,\,\, \Rightarrow
    \rm{pulse \,\,\, solutions},\\
  \\
G\left( \eta_1 \right)\,-\,G\left( \eta_2 \right)>0,\,\,\,\,\,\,\,\,\, \text{for} \,\,\,\,\,2<s<3, \,\,\,\,\,\,\,\,\, \Rightarrow \rm{anti-pulse \,\,\, solutions}.\\
\end{array}} \right.
\label{app_8}
\end{equation}

\begin{figure}[h!]
\centering
\includegraphics[width=2.0in,angle=0]{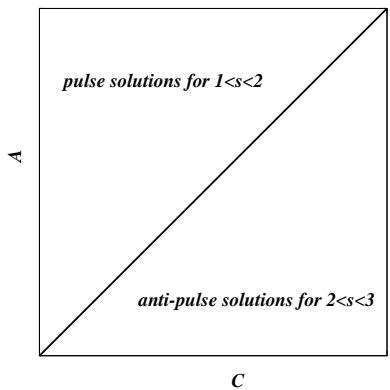}
\caption{Representation of the phase diagram in the $(A,C)$ plane
  with: {\em i)} for $1<s<2$ anti-kink solutions for the $G$,
  the corresponding counterpart for the field $u$ being pulse solutions;
  {\em ii)} for $2<s<3$ kink solutions for the $G$, corresponding to
  anti-pulse solutions for the field $u$.}
  \label{phase diagramme 1}
\end{figure}

The corresponding behaviour of the solutions of the FBE is represented in
Fig.~\ref{phase diagramme 1}, where we confine ourself to the case $C>0$ with
$A>0$ (for $C<0$, since $A>0$ one has anti-pulse solutions
for $1<s<2$). The results are summarized by concluding
for any value of $s$ in the range $1<s<2$ one obtains
    pulse solutions, while in the range $2<s<3$, one obtains
    anti-pulse solutions.
\vspace{0.25cm}

\section{Conclusions}
\label{conclusions}

We have introduced a method which, relying on an appropriate ansatz, allows us to obtain exact solutions of the fractional differential
equation derived in the continuum limit from the FPU lattice with long-range interactions.
We are able to provide solutions of the effective Fractional Bousinessq Equation (FBE) which depend on the parameters of the model,
in particular on the exponent $s$. Two classes of solutions emerge depending on the value of $s$: {\it i)} {\it pulse} solutions
displaying a ``bump'' over a constant field in the range $1<s<2$; {\it ii)} {\it anti-pulse} solutions
showing a ``hole'' in a constant field in the range $2<s<3$. As time progresses, the analytical solution reveals that both the ``bump'' and the
``hole'' damp down and, asymptotically at infinite time, the field becomes constant in space $x$, showing the delocalization
of the initially localized condition. We have focused our analysis on the $\alpha$-FPU model,
but the method we present could be extended to the $\beta$-FPU model as well.

On one hand this is a specific property of a certain class of solutions, but on the other this could be an indication of the fact that
the solutions of the FBE tend in general to delocalize in presence of long-range interactions. This is at variance
with the short-range case where it is well known that a wide class of initial conditions leads to the creation of trains of solitons
\cite{Zabusky1965,pace}.

Our results crucially depends on the ansatz presented in Section \ref{ansatz}.
Numerical solutions of the lattice FPU model with long-range interactions would be needed in order to confirm
the analytical predictions. In particular, we can already expect that localized
solutions become progressively more unstable as soon as $s$ decreases. Preliminary
numerical investigation indeed do confirm that choosing a solution that is initially localized
without initial velocity, one obtains stable solutions as soon as $s>3$~\cite{pham}. However,
the ansatz (\ref{alpha77}) requires that a suitable initial velocity is
considered and to do a proper comparison with analytical results one should appropriately choose the initial
velocities $\dot{u}_i$ at time $t=0$. Work in progress on this point is currently on-going in the region
$1<s<3$, where large enough sizes of the lattice have to be considered.

\vspace{1cm}

\section*{Acknowledgements}
It is our pleasure to acknowledge helpful conversations with Matteo Gallone,
Jonathan Pham and Nathan Nkouessi Tchepemen. One of authors, G.N.B. Chendjou, acknowledges the hospitality of the Faculty of Exact and Natural Sciences of Ivane Javakhishvili Tbilisi  State University of Tbilisi, Georgia, during November 13-26, 2017, where part of the work reported here was carried out. The support from the OFID Postgraduate Fellowship Programme at ICTP and from the ICTP/IAEA Sandwich Training Educational Programme is gratefully acknowledged. The work is supported in part by travel grants from PICS CNRS (France), CNR (Italy) and SRNSF (Georgia) grant Nos 04/01
and 04/24.  

 \bigskip

\appendix

 \newpage

 \bigskip

\end{document}